\documentclass[a4paper,conference]{IEEEtran}

\usepackage{cite}      
\usepackage{graphicx}  
\usepackage{subfigure} 
\usepackage{stfloats}  
\usepackage{amsmath,amssymb}   
\usepackage{multirow,bigstrut}

\usepackage{fancyheadings}
\begin{document}

\title{A Switch for Artificial Resistivity and Other Dissipation Terms}

\author{\IEEEauthorblockN{Terrence S. Tricco and Daniel J. Price}
\IEEEauthorblockA{Monash Centre for Astrophysics\\
Monash University\\
Melbourne, Vic, 3800,
Australia\\
terrence.tricco@monash.edu, daniel.price@monash.edu}
}

\maketitle

\begin{abstract}
We describe a new switch to reduce dissipation from artificial resistivity in Smoothed Particle Magnetohydrodynamics simulations.  The switch utilises the gradient of the magnetic field to detect shocks, setting $\alpha_B = h \vert \nabla {\bf B} \vert / \vert {\bf B} \vert$.  This measures the relative degree of discontinuity, and the switch is not dependent on the absolute field strength.  We present results comparing the new resistivity switch to the switch of Price \& Monaghan (2005), showing that it is more robust in capturing shocks (especially in weak fields), while leading to less overall dissipation.  The design of this switch is generalised to create similar switches for artificial viscosity and thermal conduction, with proof of concept tests conducted on a Sod shock tube and Kelvin-Helmholtz instabilities.
\end{abstract}

\section{Introduction}

Artificial resistivity \cite{pm05} is included in Smoothed Particle Magnetohydrodynamics (SPMHD) \cite{pm04a, pm04b, pm05} simulations to capture shocks and discontinuities in the magnetic field, similar to the use of artificial viscosity for hydrodynamic shocks.  This is accomplished by dissipating the magnetic field about the discontinuity so that the pre- and post-shock states are represented correctly.  Similar techniques exist for the treatment of contact discontinuities \cite{monaghan97} and interfacial flows \cite{molteni&colagrossi09}.  

This dissipation is unnecessary away from discontinuities.  For astrophysical simulations where high magnetic (and kinetic) Reynold's numbers are common (e.g., $\gtrsim 10^6$ for the interstellar medium), minimising dissipation is critical.   Therefore it is important to apply artificial resistivity in a targeted manner, switching on dissipation only near shocks.

Such switches exist for artificial viscosity.  The method of Morris and Monaghan \cite{mm97} use $-\nabla \cdot {\bf v}$ as a shock indicator, switching on artificial viscosity in regions of convergent flow.  Recently, Cullen and Dehnen \cite{cd10} have designed a new switch based on using ${\rm d} (\nabla \cdot {\bf v}) / {\rm d}t$ as the shock indicator, which they found detects shocks earlier while providing less overall dissipation.  Read and Hayfield \cite{read&hayfield2012} have a similar switch that uses $\nabla (\nabla \cdot {\bf v})$.

A switch for artificial resistivity was suggested by Price and Monaghan \cite{pm05} (henceforth PM05) analogous to the Morris and Monaghan \cite{mm97} viscosity switch.  However, Price, Tricco, and Bate \cite{ptb12} found even when using this switch, unwanted dissipation was still large enough to suppress the formation of protostellar jets in simulations of star formation.  As will be shown in Sec.~\ref{sec:mhdturb}, this switch also fails to capture shocks in very weak magnetic fields which would be important in simulations of cosmological and galaxy-scale magnetic fields.

In this paper, we present a new switch \cite{tp13} for artificial resistivity that robustly detects shocks and discontinuities with less overall dissipation than the PM05 switch.  We also generalise the concept to artificial viscosity and thermal conduction.  We begin with a general discussion in Sec.~\ref{sec:spmhd} to introduce the equations of SPMHD and the dissipation equations for artificial viscosity, resistivity, and thermal conduction.  The new artificial resistivity switch is introduced in Sec.~\ref{sec:newresis}, with switches of similar design constructed for artificial viscosity (Sec.~\ref{sec:newvisc}) and thermal conduction (Sec.~\ref{sec:newcond}).  Testing of the new switch is performed in Sec.~\ref{sec:resistests}, focusing on correctness of shock results, robustness of shock detection, and ability to minimise dissipation.  The other dissipation switches are explored in proof of concept tests in Sec.~\ref{sec:visctest} and Sec.~\ref{sec:condtest}.  Conclusions are drawn in Sec.~\ref{sec:conclusion}.

\section{Smoothed Particle Magnetohydrodynamics}
\label{sec:spmhd}
Ideal magnetohydrodynamics (MHD) is the coupling of the Euler equations with Maxwell's equations of electromagnetism under the assumption of a perfectly conducting fluid (i.e., no Ohmic resistance).  This yields the familiar set of Euler equations with a contribution in the momentum equation from the Lorentz force and an induction equation to describe the evolution of the magnetic field.  

The SPMHD equations solved are
\begin{align}
\rho_a &=  \sum_b m_b W_{ab} (h_a), \hspace{8mm} h_{a} = h_{\rm fac} \left( \frac{m_{a}}{\rho_{a}}\right)^{1/n_\text{dim}}, \label{eq:sphcty} \\
\frac{{\rm d}{\bf{v}}_a}{{\rm d}t} &= 
- \sum_b m_b \left[\frac{P_a}{\Omega_a \rho_a^2} \nabla_a W_{ab}(h_a) + \frac{P_b}{\Omega_b \rho_b^2} \nabla_b W_{ab}(h_b) \right] \nonumber \\
&+ \sum_b m_b \left[\frac{{\bf M}_a}{\Omega_a \rho_a^2}\cdot \nabla_a W_{ab}(h_a) + \frac{{\bf M}_{b}}{\Omega_b \rho_b^2} \cdot \nabla_b W_{ab}(h_b) \right], \label{eq:spmhd-momentum-eqn} \\
 \frac{{\rm d}{\bf{B}}_a}{{\rm d}t} &= - \frac{1}{\Omega_a \rho_a} \sum_b m_b \bigg[ {\bf{v}}_{ab} \left( {\bf{B}}_a \cdot \nabla_a W_{ab}(h_a) \right) \nonumber \\
& \hspace{28mm}- {\bf{B}}_a \left( {\bf{v}}_{ab} \cdot \nabla_a W_{ab}(h_a) \right) \bigg], \label{eq:sphind}
\end{align}
where ${\bf v}$ is the velocity and ${\bf B}$ is the magnetic field.  The density, $\rho$, and smoothing length, $h$, are self-consistently derived through iteration of (\ref{eq:sphcty}), with $n_\text{dim}$ corresponding to the number of dimensions and $h_{\rm fac} = 1.2$ relating the smoothing length to the local particle spacing.  Variable smoothing length gradients are handled by $\Omega$ (see \cite{sh02}).  

The thermal pressure, $P$, is obtained through a suitable equation of state.  The momentum equation (\ref{eq:spmhd-momentum-eqn}) is derived from the Lagrangian, producing the usual thermal pressure terms with the Lorentz force expressed in terms of the Maxwell stress tensor,
\begin{equation}
{\bf M} = \frac{{\bf B}{\bf B}}{\mu_0} - \frac{B^2}{2\mu_0} {\bf I} .
\label{eq:divM}
\end{equation}

The induction equation (\ref{eq:sphind}) is derived from $\partial {\bf B}/\partial t = \nabla \times ({\bf v} \times {\bf B})$.  However, $\nabla \cdot {\bf B} = 0$ from Maxwell's equations, and consequently terms containing $\nabla \cdot {\bf B}$ are subtracted from the SPMHD momentum equation introducing a small amount of non-conservation of energy and momentum, but greatly enhancing stability and performance.  The constrained hyperbolic divergence cleaning algorithm of Tricco and Price \cite{tp12} is used to minimise divergence error.

Artificial viscosity and resistivity terms are added to the momentum and induction equations in order to capture hydrodynamic and magnetic shocks.  Thermal conduction is added when integrating an energy equation to handle contact discontinuities.  These dissipation terms are necessary, as the differential SPH equations assumes differentiability leading to errors at the discontinuity.  By smoothing the discontinuity over several smoothing lengths, these errors are removed.

\subsection{Artificial Viscosity and Thermal Conduction}
\label{sec:viscosity}

We use the artificial viscosity from Monaghan \cite{monaghan97} formulated by analogy with Riemann solvers, giving
\begin{equation}
 \left( \frac{{\rm d}{\bf v}_a}{{\rm d}t} \right)_\text{diss} = \sum_b m_b \frac{\alpha v_\text{sig}}{\overline{\rho}_{ab}} ({\bf v}_a - {\bf v}_b) \cdot \hat{{\bf r}}_{ab} \nabla_a W_{ab} ,
\label{eq:visc}
\end{equation}
where $\alpha$ is a dimensionless parameter of order unity.  The signal velocity corresponds to the speed of information between the two particles,
\begin{equation}
 v_\text{sig} = 0.5 \left( c_a + c_b - \beta {\bf v}_{ab} \cdot \hat{{\bf r}}_{ab} \right) ,
\label{eq:vsig}
\end{equation}
with the $\beta=2$ term correcting for the relative motion of the particles preventing particle interpenetration through the shock.  For pure hydrodynamic flows, information propagates at the speed of sound, $c$.  For magnetohydrodynamics, the fast MHD wave speed,
\begin{equation}
 v = \frac{1}{\sqrt{2}} \left[ \left(c^2 + v_A^2\right) + \left[ (c^2 + v_A^2)^2 - 4 c^2 v_A^2 (\hat{{\bf B}} \cdot \hat{{\bf r}}_{ij}) \right]^{1/2} \right]^{1/2} .
\label{eq:vsigfastmhd}
\end{equation}
is used with $v_A = B/ \sqrt{\mu_0 \rho}$ corresponding to the Alfv\'en speed.  

A thermal conduction term of the form
\begin{equation}
 \left( \frac{{\rm d}u}{{\rm d}t} \right)_\text{cond} = - \sum_b m_b \frac{\alpha_u v_\text{sig}^u}{\overline{\rho}_{ab}} (u_a - u_b) \hat{{\bf r}}_{ab} \nabla_a W_{ab} 
\end{equation}
also originates from Monaghan \cite{monaghan97}.  Price \cite{price08} suggested using
\begin{equation}
 v_\text{sig}^u = \sqrt{\frac{\vert P_a - P_b \vert}{\overline{\rho}_{ab}}},
\end{equation}
which is adopted in this work.

Morris and Monaghan \cite{mm97} developed a switch to decrease viscous dissipation in regions away from discontinuities.  This switch lets $\alpha$ be individual to each particle, which is integrated according to
\begin{equation}
 \frac{{\rm d} \alpha_a}{{\rm d}t} = \max(- \nabla \cdot {\bf v}_a, 0) - \frac{\alpha_a - \alpha_\text{min}}{\tau} .
\label{eq:intalpha}
\end{equation}
A range of $\alpha \in [0.1, 1]$ is enforced.  Equation (\ref{eq:intalpha}) uses a source term to increase $\alpha$ when shocks are detected ($\nabla \cdot {\bf v} < 0$), along with a decay term to reduce $\alpha$ over time.  The decay timescale, $\tau = h / C c$, is chosen so that viscosity persists for approximately five smoothing lengths post-shock ($C\sim0.1$).

Cullen and Dehnen \cite{cd10} improved on this switch by defining a new shock indicator, 
\begin{equation}
 A_a = \xi \max\left(-\frac{{\rm d}( \nabla \cdot {\bf v}_a)}{{\rm d}t}, 0\right),
\end{equation}
where $\xi$ is a limiter for shear flows similar to the Balsara limiter \cite{balsara95}.  In their work, $\alpha$ is set equal to 
\begin{equation}
 \alpha_a = \alpha_\text{max} \frac{h_a^2 A_a}{v_\text{sig}^2 + h_a^2 A_a}
\end{equation}
whenever it exceeds the current value, otherwise is decayed slowly using
\begin{equation}
 \frac{{\rm d}\alpha_a}{{\rm d}t} = - \frac{\alpha_a - \alpha_\text{min}}{\tau} .
\end{equation}
They use an improved estimator for $\nabla \cdot {\bf v}$ and ${\rm d} (\nabla \cdot {\bf v}) / ({\rm d} t)$, and find their method allows one to use $\alpha_\text{min}=0$.

\subsection{Artificial Resistivity and the PM05 Switch}

The standard implementation of artificial resistivity \cite{pm05} adds terms to the induction equation (\ref{eq:sphind}) of the form,
\begin{equation}
 \left( \frac{{\rm d}{\bf B}_a}{{\rm d}t} \right)_\text{diss} = \rho_a \sum_b m_b \frac{\alpha_B v^{B}_\text{sig}}{\overline{\rho}_{ab}^2} ({\bf B}_a - {\bf B}_b) \hat{\bf{r}}_{ab} \cdot \nabla_a W_{ab} .
\label{eq:resistivity}
\end{equation}
As in (\ref{eq:visc}), $\alpha_B$ is a dimensionless parameter of order unity.

The signal velocity for artificial resistivity is not straightforward as there are three wave solutions in ideal magnetohydrodynamics: fast and slow waves, which are composite sound and magnetic waves, and Alfv\'en waves which are purely magnetic.  From (\ref{eq:vsigfastmhd}), it can be seen that the Alfv\'en and fast wave speeds will be comparable for strong magnetic fields, but for weak fields may differ significantly.  Since it is not possible to determine the type of shock without reconstructing the full Riemann state, we choose to use the fast wave speed.  Furthermore, the $\beta$ term in (\ref{eq:vsig}) is not used as it is unnecessary to worry about particle interpenetration and otherwise causes excessive dissipation.

PM05 introduced a switch for artificial resistivity where $\alpha_B$ is integrated according to
\begin{equation}
\label{eq:intalphab}
 \frac{{\rm d}\alpha_{B,a}}{{\rm d}t} = \max\left( \frac{\vert \nabla \cdot {\bf B}_a \vert}{\sqrt{\mu_{0}\rho}}, \frac{\vert \nabla \times {\bf B}_a \vert}{\sqrt{\mu_{0}\rho}}\right) - \frac{\alpha_{B,a}}{\tau} 
\end{equation}
with $\alpha_B \in [0,1]$.  This switch is analogous to the Morris and Monaghan \cite{mm97} switch for artificial viscosity (\ref{eq:intalpha}), with the source terms produced from dimensional analysis.

\section{New Artificial Resistivity Switch}
\label{sec:newresis}

Our new switch is constructed using the full gradient matrix of the magnetic field, $\nabla {\bf B}$, as the shock indicator.  Rather than use a time integrated value for $\alpha_B$, instead at each time step it is directly set to
\begin{equation}
 \alpha_{B,a} = \frac{h_a \vert \nabla {\bf B}_a \vert}{\vert {\bf B}_a \vert} 
\label{eq:newalpharesis}
\end{equation}
within the range $\alpha_B \in [0,1]$.  One advantage to this approach is that there is no time delay when increasing $\alpha_B$.  

Another nice property of this switch is that by normalising $\nabla {\bf B}$ by the magnitude of the magnetic field, dependence on ${\bf B}$ is removed.  This yields a measure of the relative degree of discontinuity in the field, naturally producing values of $\alpha_B$ in the desired range allowing for shocks to be detected at all field strengths.  Negligible $\alpha_B$ values are produced away from shocks.  Since $v_A \propto B$, this leads to a quantity which is related to the Alfv\'enic Mach number.  

The full gradient of the magnetic field is calculated using a standard first derivative SPH operator (e.g., \cite{price12}),
\begin{equation}
\nabla{\bf B}_{a} \equiv \frac{\partial B^i_{a}}{\partial x^j_{a}} \approx -\frac{1}{\Omega_a \rho_a} \sum_b m_b ({B}^{i}_a - {B}^{i}_b) \nabla_a^j W_{ab}(h_{a}) .
\end{equation}
The curl of the magnetic field was tested as the shock indicator in place of $\nabla {\bf B}$, but found that it did not perform as well for complicated shock interactions.  The norm of the matrix is calculated using the 2-norm,
\begin{equation}
\vert \nabla {\bf B} \vert \equiv \sqrt{ \sum_i \sum_j \left\vert  \frac{\partial B^i_{a}}{\partial x^j_{a}} \right\vert^2 } .
\end{equation}
Several other choices for norms were considered without finding significant differences.

\subsection{Generalisation to Other Dissipation Terms}

The design concept of the new artificial resistivity switch could be extended to other dissipation terms.  In the following, we construct new switches for artificial viscosity and thermal conduction based on the general idea of a normalised shock indicator.  

\subsubsection{New Artificial Viscosity Switch}
\label{sec:newvisc}

For the artificial viscosity switch, we continue to use $- \nabla \cdot {\bf v}$ as the shock indicator, as in (\ref{eq:intalpha}).  It would be unwise to use $\vert {\bf v} \vert$ for the normalisation as this would break Galilean invariance.  Instead, the sound speed is used, relating the quantity to the Mach number.

It is also important that artificial viscosity is applied to the wake of the shock to reduce post-shock oscillations of the particles.  Therefore, $\alpha$ is reduced over time using an integrated decay term like in (\ref{eq:intalpha}).  

The resulting switch is therefore to set 
\begin{equation}
 \alpha_a = - \frac{h_a \nabla \cdot {\bf v}_a}{c} 
\label{eq:newalphavisc}
\end{equation}
when greater than the current value of $\alpha_a$, otherwise $\alpha_a$ is reduced on the next time step according to
\begin{equation}
 \frac{{\rm d}\alpha_a}{{\rm d}t} = -\frac{\alpha_a}{\tau} ,
\end{equation}
where $\tau$ has the same meaning as in (\ref{eq:intalpha}).  By following the considerations outlined above, this viscosity switch is quite similar in principle to the switch of Cullen and Dehnen \cite{cd10}, albeit with a simpler version for (\ref{eq:newalphavisc}).

\subsubsection{New Thermal Conduction Switch}
\label{sec:newcond}

A switch for thermal conduction can be constructed by analogy to (\ref{eq:newalpharesis}).  The gradient of thermal energy is chosen to detect discontinuities, setting
\begin{equation}
 \alpha_{u,a} = \frac{h_a \vert \nabla u_a \vert}{\vert u_a \vert} .
\end{equation}

\section{Artificial Resistivity Switch Tests}
\label{sec:resistests}

The new switch for artificial resistivity is tested for three criteria: i) correctness of shock profiles, ii) ability to minimise dissipation away from shocks, and iii) robustness of shock detection.  A three-dimensional MHD shock tube which produces three different types of magnetic shocks is used to verify the first criterion.  The level of dissipation between the new switch and the PM05 switch is compared using the Orszag-Tang vortex.  Finally, a weak magnetic field in the presence of Mach 10 turbulence is used to gauge the robustness of the switch.

\subsection{3D MHD Shock Tube}

The shock tube problem from Dai and Woodward \cite{dw94} is used to test the ability of the switch to capture magnetic shocks.  It has three-dimensional velocity and magnetic field structure producing seven shocks: fast shocks, slow shocks, and rotational discontinuities travelling in either direction, with a contact discontinuity in the centre.  The initial left state ($x<0$) is ($\rho$, $P$, $v_x$, $v_y$, $v_z$, $B_y$) $=$ (1.08, 0.95, 1.2, 0.01, 0.5, $3.6/\sqrt{4 \pi}$) and right state ($\rho$, P, $v_x$, $v_y$, $v_z$, $B_y$) $=$ (1, 1, 0, 0, 0, $4/\sqrt{4 \pi}$) with $B_x=B_z=2/\sqrt{4 \pi}$ and $\gamma=5/3$.  

\begin{figure}
 \includegraphics[width=1.0\linewidth]{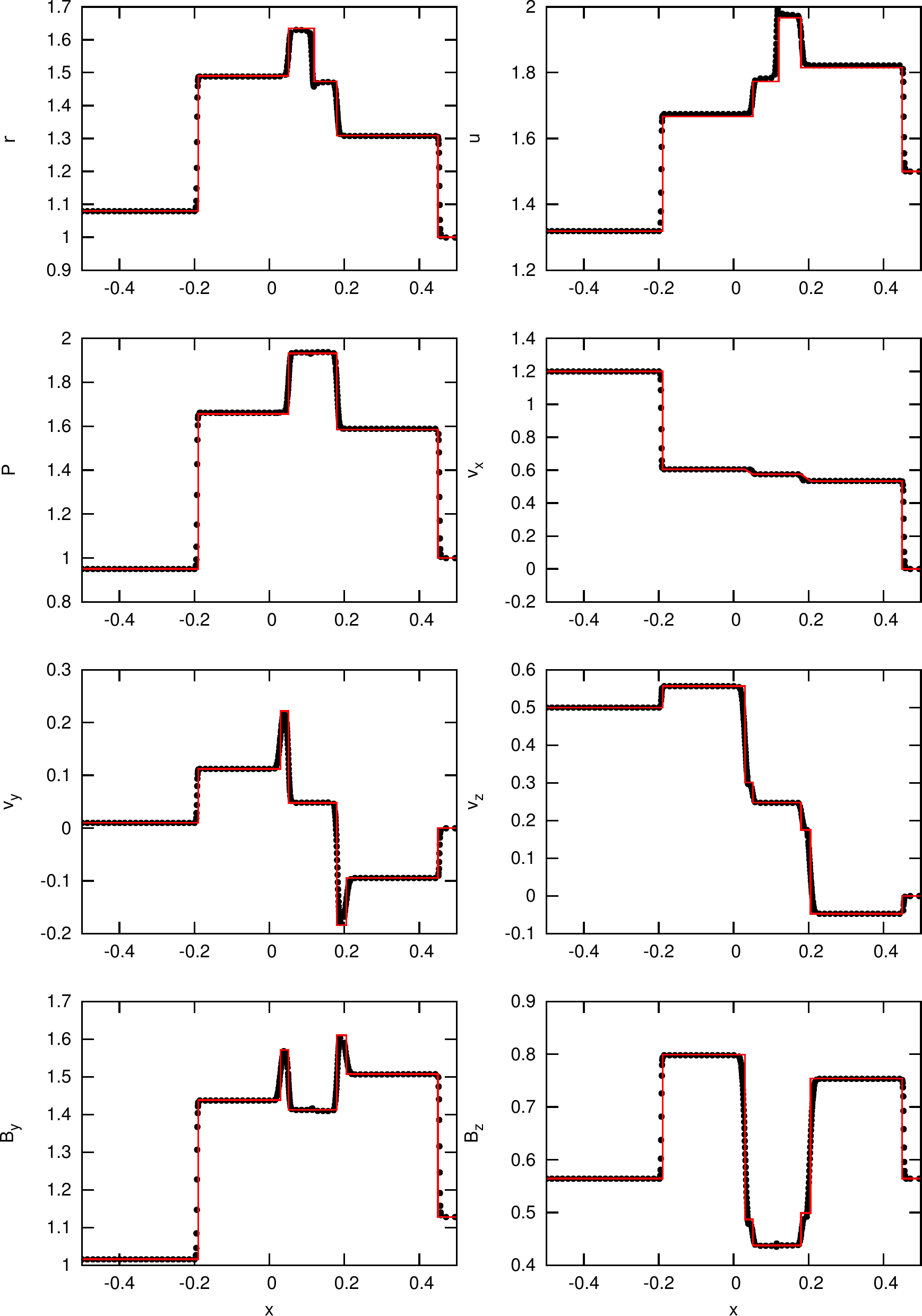}
\caption{3D MHD shock tube test with left state ($\rho$, $P$, $v_x$, $v_y$, $v_z$, $B_y$) $=$ (1.08, 0.95, 1.2, 0.01, 0.5, $3.6/\sqrt{4 \pi}$) and right state ($\rho$, P, $v_x$, $v_y$, $v_z$, $B_y$) $=$ (1, 1, 0, 0, 0, $4/\sqrt{4 \pi}$) with $B_x=B_z=2/\sqrt{4 \pi}$ at $t=0.2$. Black circles are the particles and the red line is the solution from Ryu and Jones \cite{rj95}.  Excellent agreement is noted in all shock profiles.}
\label{fig:shock2a}
\end{figure}

The shock tube is performed in 3D using 955$\times$12$\times$12 particles for the left state and 597$\times$12$\times$12 particles for the right state initially arranged on cubic lattices.  Fixed artificial viscosity ($\alpha=1$) and thermal conduction ($\alpha_u=0.5$) has been used, along with the quintic spline to minimise noise from the particles remeshing in the $y$ and $z$ directions. Results at $t=0.2$ are presented in Fig.~\ref{fig:shock2a} along with a reference solution from Ryu and Jones \cite{rj95}.  Excellent agreement is observed between the two solutions.

\subsection{Orszag-Tang Vortex}

The Orszag-Tang vortex \cite{ot79} is a widely used test problem for MHD codes.  It is a two-dimensional problem consisting of a velocity vortex overlaid with a magnetic vortex that develops into magnetic turbulence.  Here we use it to compare the level of unnecessary dissipation between our switch and the PM05 switch.  The initial state has $\rho=25/(36\pi)$, $P=5/(12\pi)$, with $\gamma=5/3$.  The velocity vortex is given by ${\bf v}=[-\sin(2\pi y), \sin(2\pi x)]$ and magnetic vortex by ${\bf B}=[-\sin(2\pi y), \sin(4\pi x)]$.

Due to the non-linear behaviour at late times, the test performed to $t=1$, which encompasses the linear evolution phase before turbulence fully develops.  In the absence of an analytic solution, the test is compared for three different resolutions of $256^2$, $512^2$, and $1024^2$.  The artificial viscosity switch from (\ref{eq:intalpha}) is used, with fixed $\alpha_u=0.1$ thermal conduction.

\begin{figure}
 \setlength{\tabcolsep}{0.002\textwidth}
\begin{tabular}{ccl}
 \scriptsize{PM05 switch} & \scriptsize{New switch} 
\\
   \includegraphics[height=0.445\linewidth]{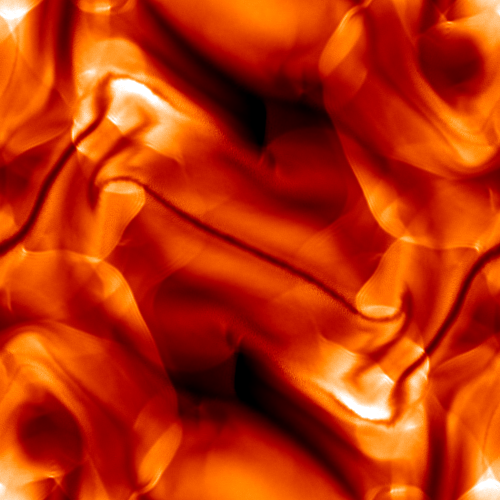} 
 & \includegraphics[height=0.445\linewidth]{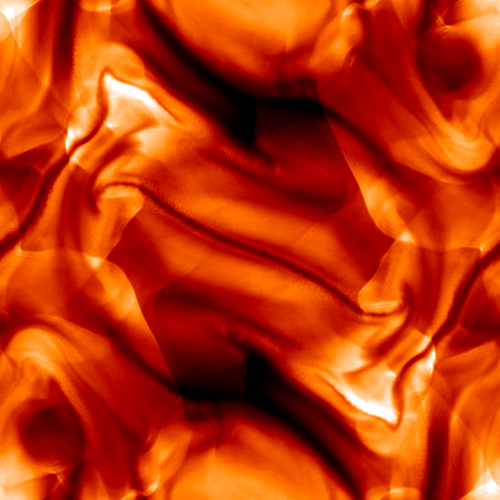} 
 & \includegraphics[height=0.445\linewidth]{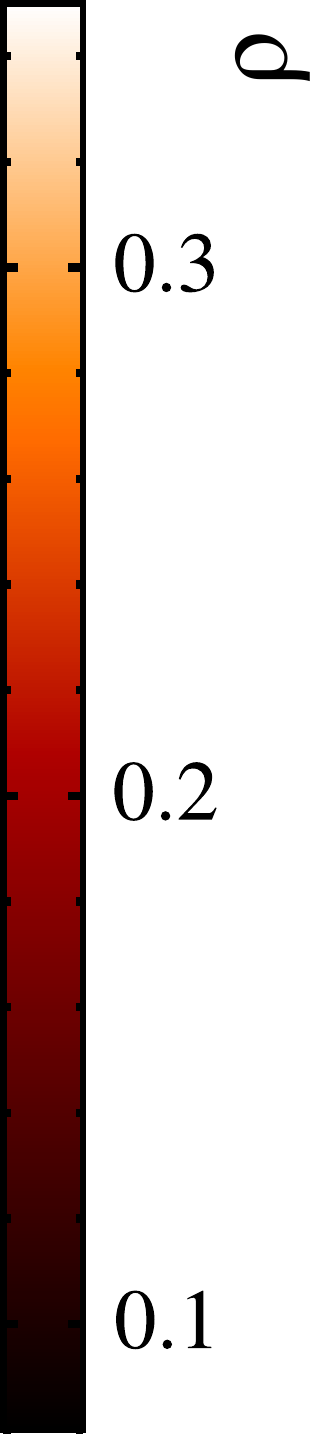}
\\
   \includegraphics[height=0.445\linewidth]{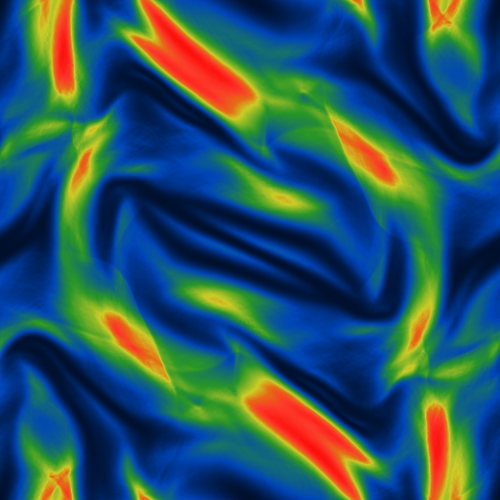}
 & \includegraphics[height=0.445\linewidth]{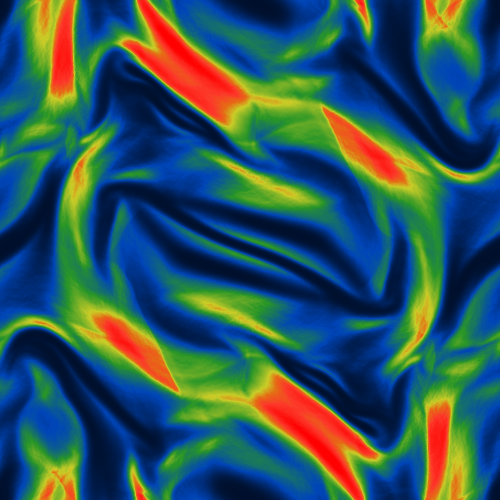}
 & \includegraphics[height=0.445\linewidth]{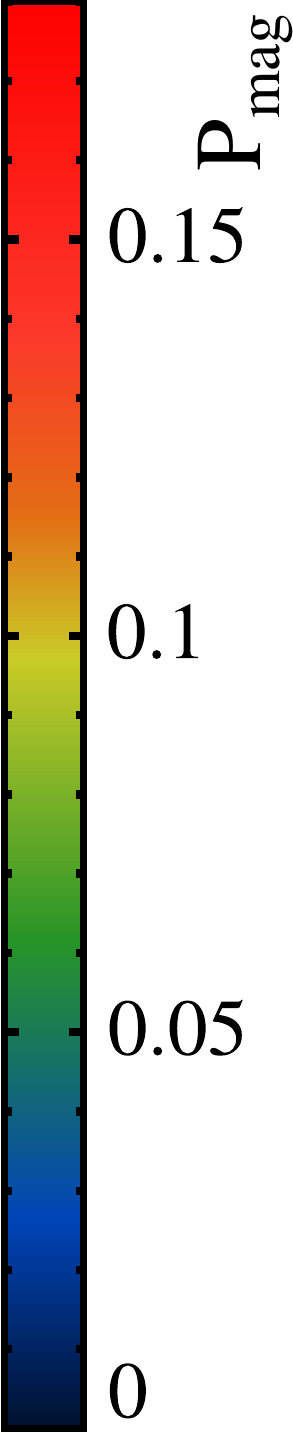}
\\
   \includegraphics[height=0.445\linewidth]{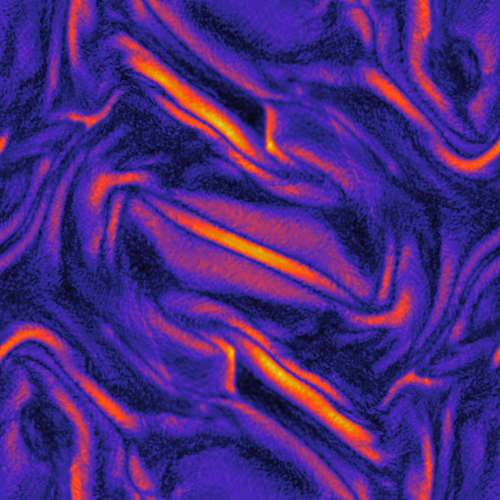}
 & \includegraphics[height=0.445\linewidth]{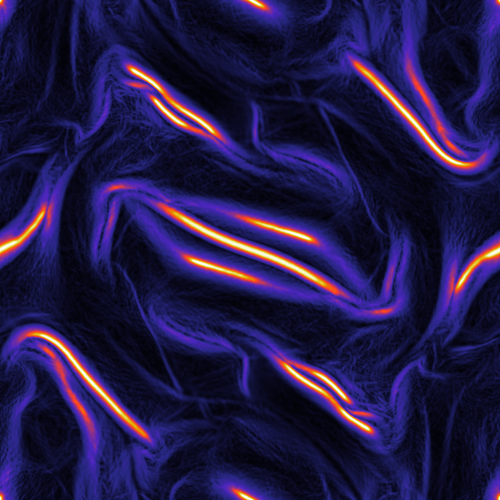}
 & \includegraphics[height=0.445\linewidth]{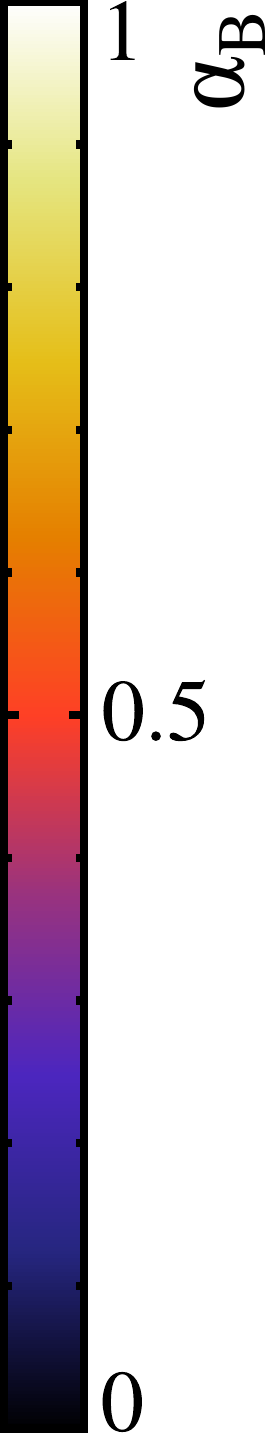}
\\
\end{tabular}
\caption{The density (top), magnetic pressure (middle), and $\alpha_B$ (bottom) of the Orszag-Tang vortex at $t=1$ for the old (left) and new (right) resistivity switches.  The new switch effectively traces the shock lines, with little dissipation present between shocks.  Some of the low density regions are more sharply defined using the new switch due to the lower smoothing of magnetic field structure.}
\label{fig:orszag-alphab}
\end{figure}

Renderings of the density, magnetic pressure, and $\alpha_B$ in the system for the $512^2$ resolution case at $t=1$ are presented in Fig.~\ref{fig:orszag-alphab}.  There is agreement between the two switches in the density and magnetic pressure fields, however it is noticed that subtle magnetic fields, particularly in low density regions, are smoothed away with the PM05 switch.  This occurs due to the broad $\alpha_B$ regions present over shocked areas, and small, yet significant, values of $\alpha_B$ inbetween shocks. In contrast, our new switch traces the shock lines with negligible dissipation between shocks.  This is demonstrated by the mean $\alpha_B$, which is twice as high for the PM05 switch compared to our switch $(\sim0.2$ to $\sim0.1)$.


\begin{figure}
 \includegraphics[width=1.0\linewidth]{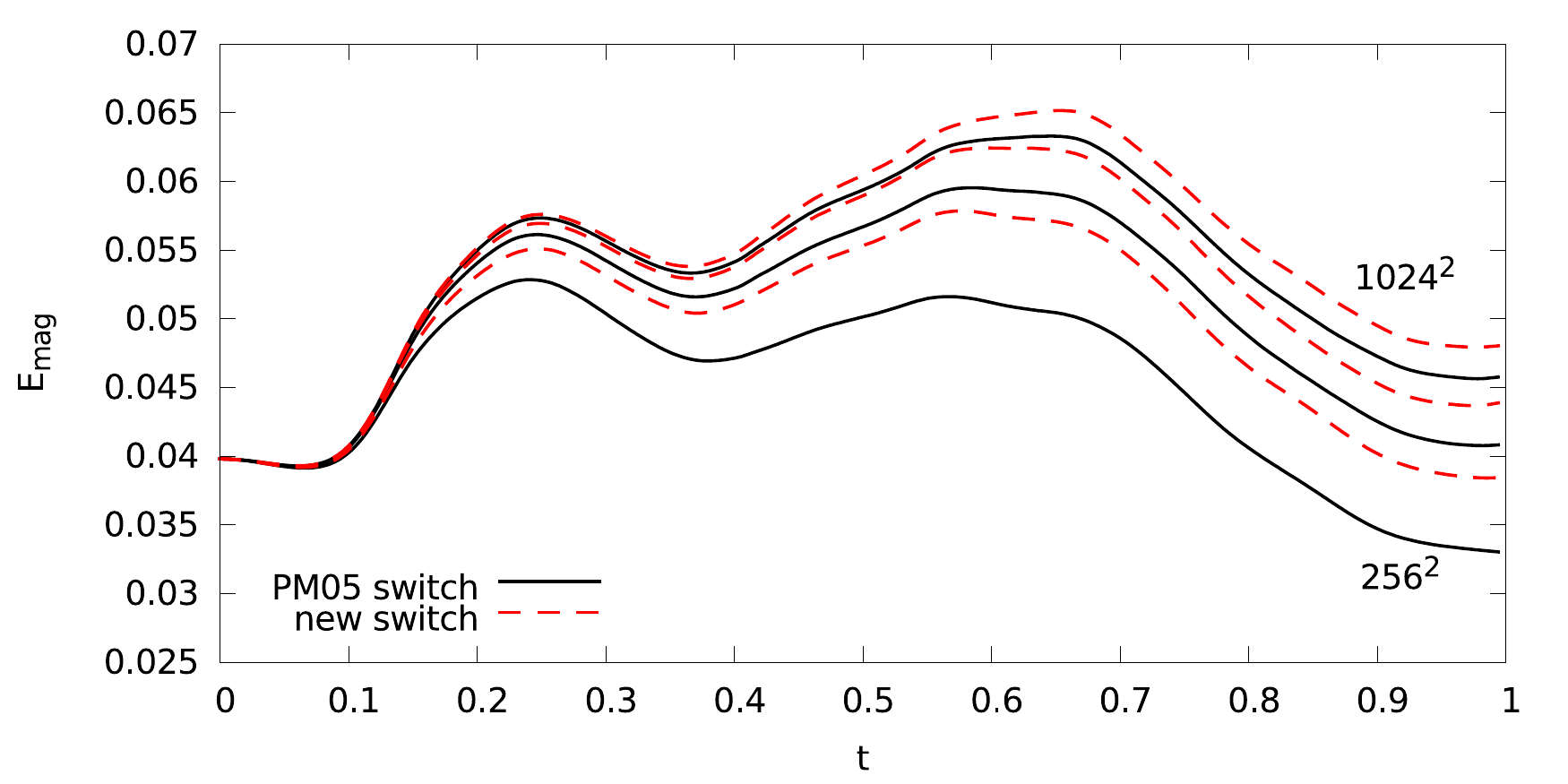}
\caption{Evolution of the magnetic energy in the Orszag-Tang vortex between the PM05 switch (solid, black lines) and new artificial resistivity switch (red, dashed lines) at resolutions of $256^2$, $512^2$, and $1024^2$.  The new switch is much less dissipative than the PM05 switch, which has similar effect to increasing the resolution.}
\label{fig:orszag-be}
\end{figure}

The evolution of magnetic energy between the two switches for the three resolutions tested in shown in Fig.~\ref{fig:orszag-be}.  It is clear that the PM05 switch dissipates more magnetic energy than the new switch.  Futhermore, the magnetic energy is converging towards larger values as the resolution is increased, thus it can be concluded that using the new switch produces an effect similar to increasing the resolution.  That is, the new switch can provide the same result as the PM05 switch at a lower resolution, which is important when dealing with large astrophysical simulations.

\subsection{Mach 10 Magnetised Turbulence}
\label{sec:mhdturb}

Our final test is of driven Mach 10 turbulence in a magnetised medium.  The initial magnetic field is extremely weak, with the magnetic energy 10 orders of magnitude smaller than kinetic energy.  Therefore, this represents a robust test of the switchs ability to detect shocks in very weak fields.  Capturing these shocks is important as they cause dynamo amplification of the magnetic field through the conversion of turbulent energy.

The initial system has $\rho=1$ and zero velocity.  An isothermal equation of state with $c=1$ is used.  A uniform magnetic field is present with $B_z = \sqrt{2} \times 10^{-5}$, such that the initial plasma $\beta$, representing the ratio of thermal pressure to magnetic pressure, is $10^{10}$.  The turbulence is driven using an Uhlenbeck-Ornstein process \cite{eswaranpope88, federrathetal10} which is a stochastic process that drives motion at low wave numbers.  The driving force is constructed in Fourier space, allowing it to be decomposed into solenoidal and compressive modes, and here we use only the solenoidal component.  This simulation mimics the pure hydrodynamic case of Mach 10 driven turbulence from Price and Federrath \cite{pricefederrath10}, but with the addition of magnetic field physics.

\begin{figure}
\setlength{\tabcolsep}{0.002\textwidth}
\begin{tabular}{ccr}
\scriptsize{PM05 switch} & \scriptsize{New switch} 
\\
   \includegraphics[height=0.438\linewidth]{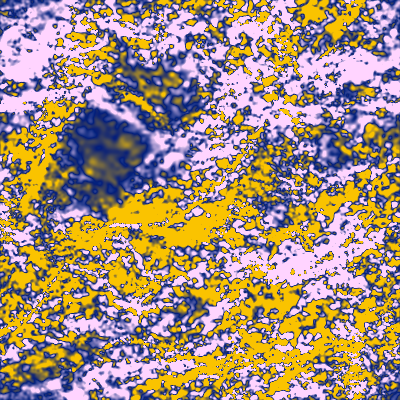}
 & \includegraphics[height=0.438\linewidth]{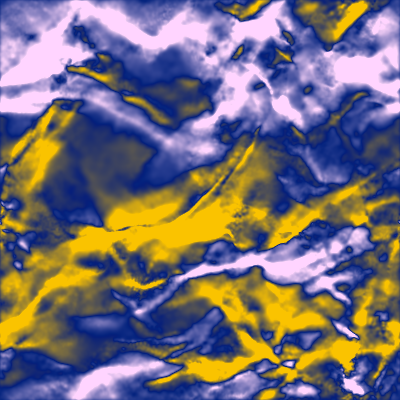}
 & \includegraphics[height=0.438\linewidth]{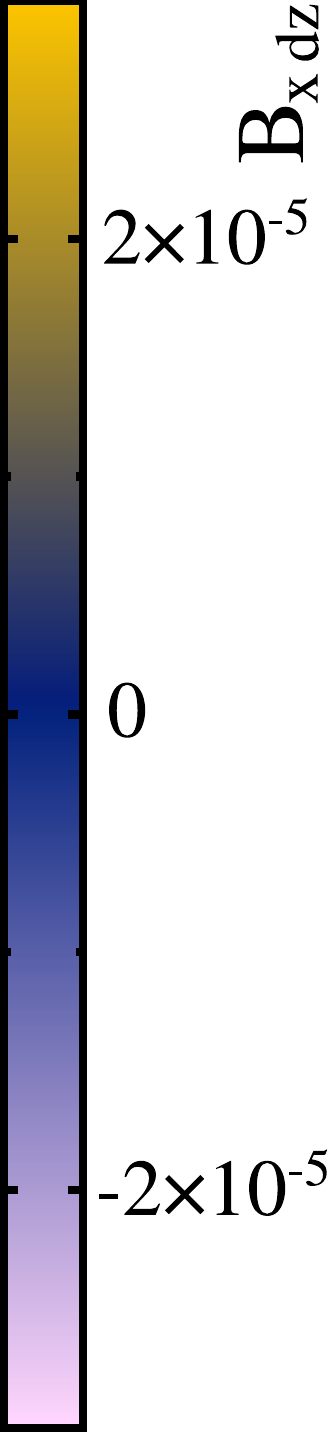}
\\
   \includegraphics[height=0.438\linewidth]{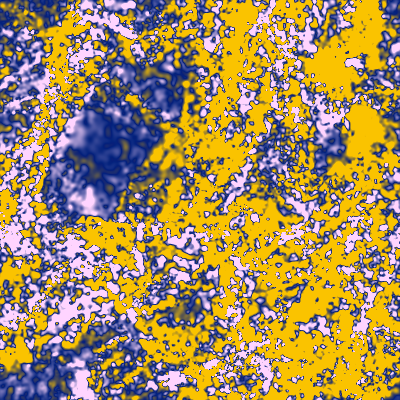}
 & \includegraphics[height=0.438\linewidth]{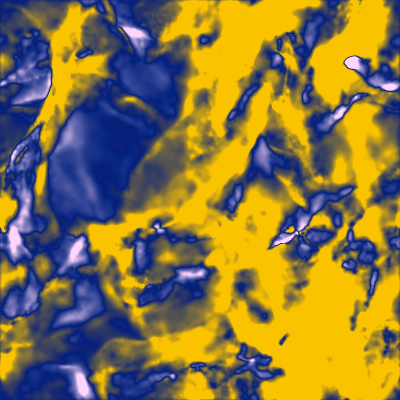}
 & \includegraphics[height=0.438\linewidth]{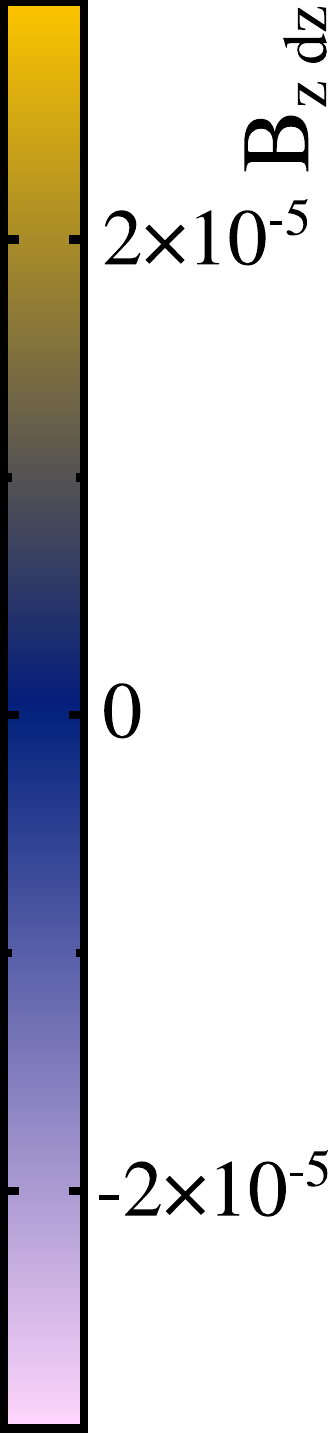}
\end{tabular}
\caption{The column integrated $x$ \& $z$ (top, bottom) magnetic field components for the PM05 (left) and new switch (right) after two turbulent turnover times (i.e., the regime of fully developed turbulence).  The magnetic field structure using the previous switch is dominated by unphysical noise due to the shocks failing to be captured (left), whereas the new switch is able to capture the shocks and the magnetic field retains its physical structure (right).}
\label{fig:mhdturb}
\end{figure}

The turbulence was simulated using $128^3$ particles using separately the PM05 switch and the new resistivity switch.  In Fig.~\ref{fig:mhdturb}, the column integrated $x$ and $z$ components of the magnetic field after two turbulent turnover times are presented.  Since the PM05 switch is proportional to ${\bf B}$, it fails to increase $\alpha_B$ to levels sufficient to treat the shocks ($\alpha_B \sim 10^{-5}$), and the shocks break apart causing unphysical noise in the magnetic field.  On the other hand, by measuring the relative degree of discontinuity in the magnetic field, the new switch is able to capture the shocks and correctly model the dynamo amplification process.

\section{Tests of Artificial Viscosity and Thermal Conduction Switches}

The efficacy of the new artificial viscosity and thermal conduction switches are examined using a standard Sod shock tube test and a test producing Kelvin-Helmholtz instabilities.  

\subsection{Viscosity: Sod Shock Tube}
\label{sec:visctest}

The Sod shock tube \cite{sod78} has become a canonical test for hydrodynamic shocks.  It consists of a fluid with a discontinuity in the density and pressure that sends a shock wave into the low density medium and a rarefaction into the high density medium, with a contact discontinuity in the centre.  Artificial viscosity is required in this test in order to treat the shock wave.  It has left state ($x<0$) $\rho=1$ and $P=1$ in contact with a fluid of $\rho=0.125$ and $P=0.1$ with $\gamma=5/3$.  Both states have zero initial motion.  

\begin{figure}
 \includegraphics[width=1.0\linewidth]{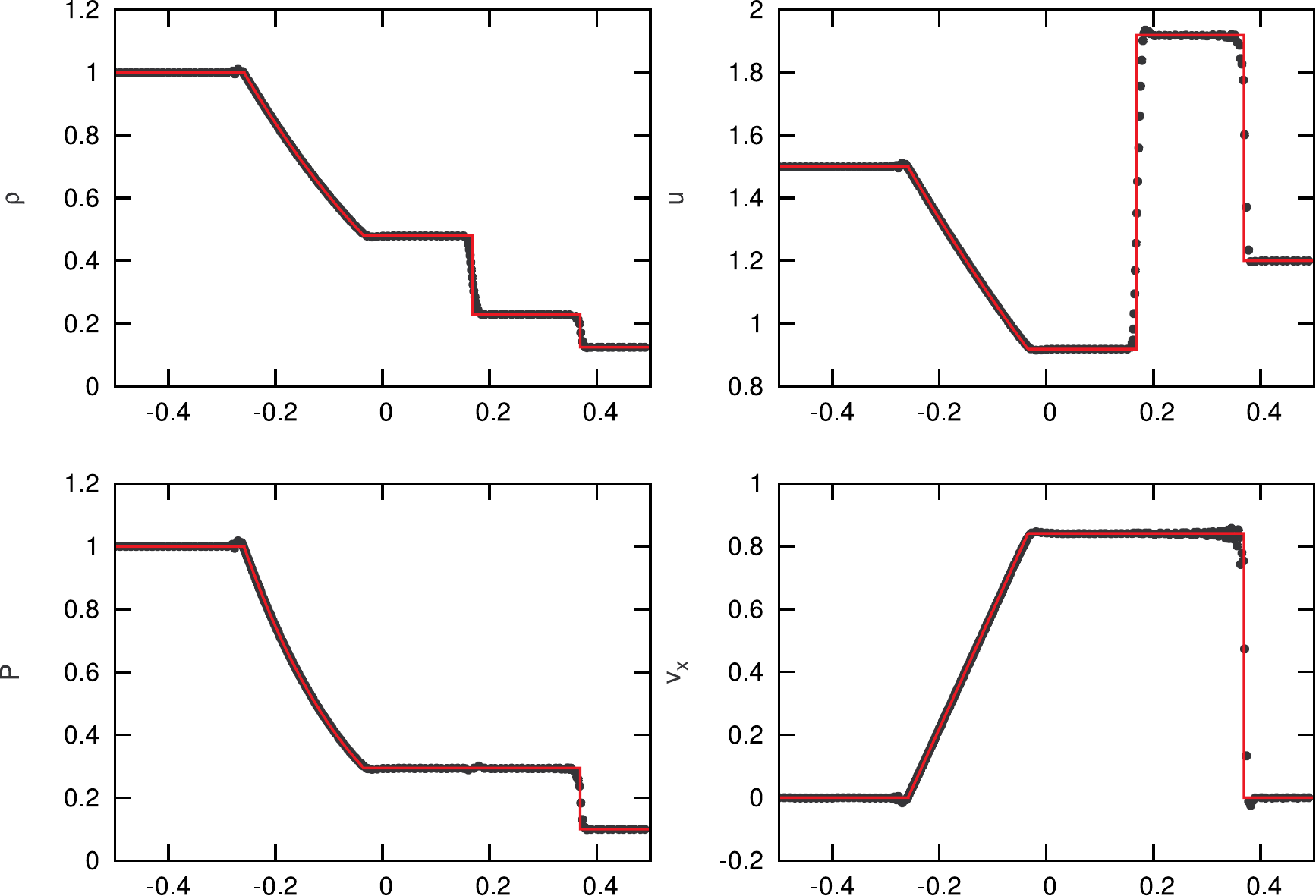}
\caption{Sod shock tube results at $t=0.2$ using the new viscosity switch described in Sec.~\ref{sec:newvisc}.  The black circle are values from the particles with the red line the Riemann solution.}
\label{fig:sodshock-results}
\end{figure}

The shock tube is simulated in 1D using $1000$ and $125$ particles for the two states, respectively.  Thermal conduction is used with fixed $\alpha_u=1$. Results at $t=0.2$ are presented in Fig.~\ref{fig:sodshock-results} along with the solution calculated from a Riemann solver.  Good agreement in all profiles is obtained between the SPH and Riemann solutions.

\subsection{Thermal Conduction: Kelvin-Helmholtz Instability}
\label{sec:condtest}

Kelvin-Helmholtz instabilities have been studied many times with SPH (e.g., \cite{price08, valckeetal10, mcnally12}).  The instability occurs when there is a velocity shear in a fluid, causing turbulence to form along the interface.  An important aspect to simulating this correctly in SPH is application of thermal conduction to treat thermal energy discontinuities across the interface.   If ignored, spurious pressure is generated preventing the fluid from mixing properly.  We use this test to investigate the ability of the new thermal conduction switch to allow mixing of the fluids across the interface, and produce the ``curls'' which are emblematic of Kelvin-Helmholtz instabilities.

The test performed here follows the initial set up of \cite{price08}.  The fluid contains two regions in a 2:1 density contrast.  The domain is $x,y=[-0.5,0.5]$ and periodic boundary conditions are used creating two interfaces along which Kelvin-Helmholtz instabilities form.  The initial density profile is
\begin{equation}
 \rho =
\begin{cases}
 2 & \text{  } \vert y \vert < 0.25, \\
 1 & \text{  } \vert y \vert > 0.25.
\end{cases}
\end{equation}
The two regions are in pressure equilibrium with uniform $P=2.5$ with $\gamma=5/3$.  The $x$-velocity is $-0.5$ for the $\rho=2$ region, and $0.5$ for the $\rho=1$ region.  The $y$-velocity is zero, however the instability is seeded with a perturbation across the interfaces by
\begin{equation}
 v_y =
\begin{cases}
 A \sin[-2 \pi (x+ 0.5) / \lambda] & +0.225 < y < +0.275, \\
 A \sin[2 \pi (x + 0.5) / \lambda] & -0.225 < y < -0.275,
\end{cases}
\end{equation}
where $A=0.025$ and $\lambda = 1/6$ giving an instability timescale of $\tau_\text{KH}=0.35$.

\begin{figure*}
 \centering
\setlength{\tabcolsep}{0.15mm}
\begin{tabular}{ccccl}
   \includegraphics[height=0.2\linewidth]{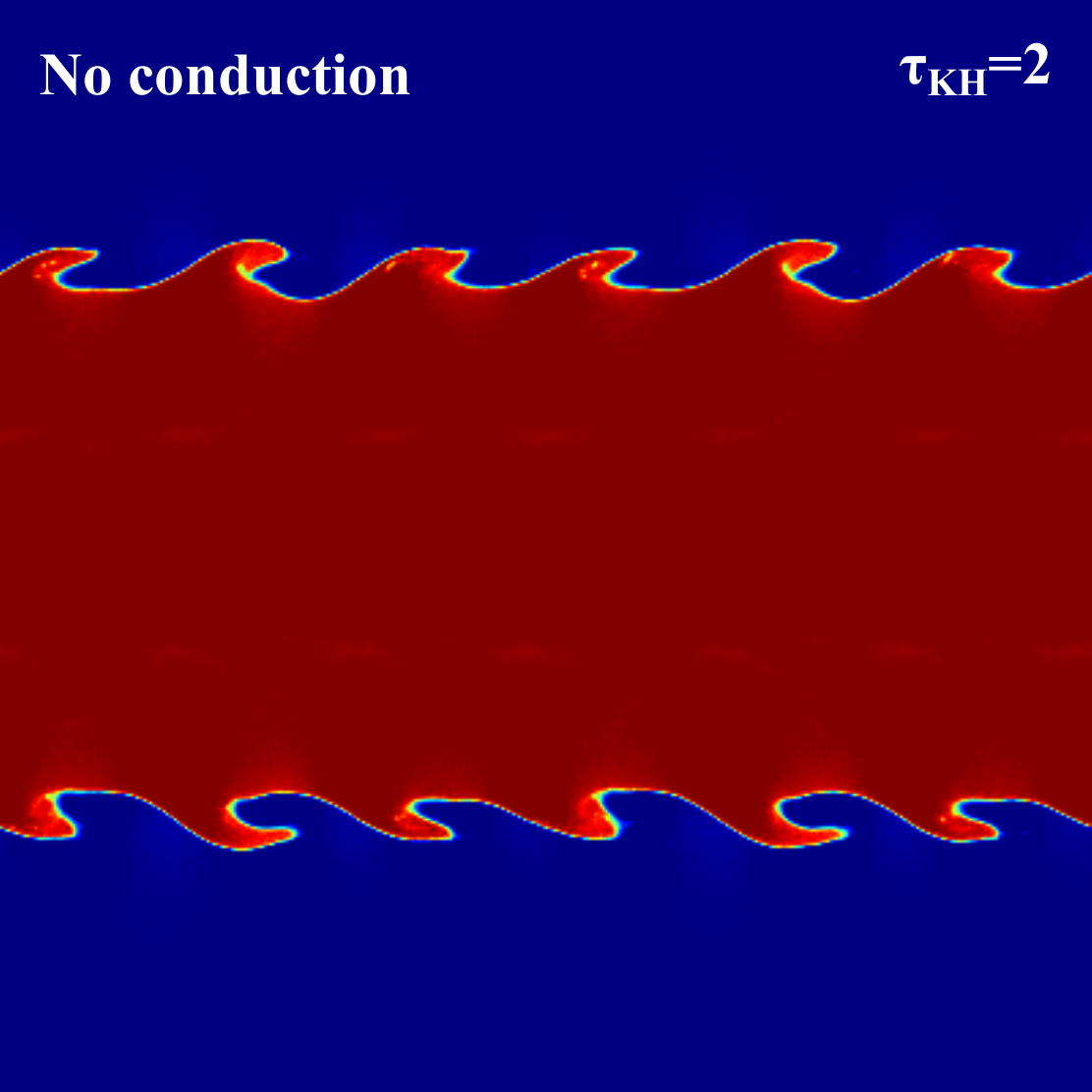}
 & \includegraphics[height=0.2\linewidth]{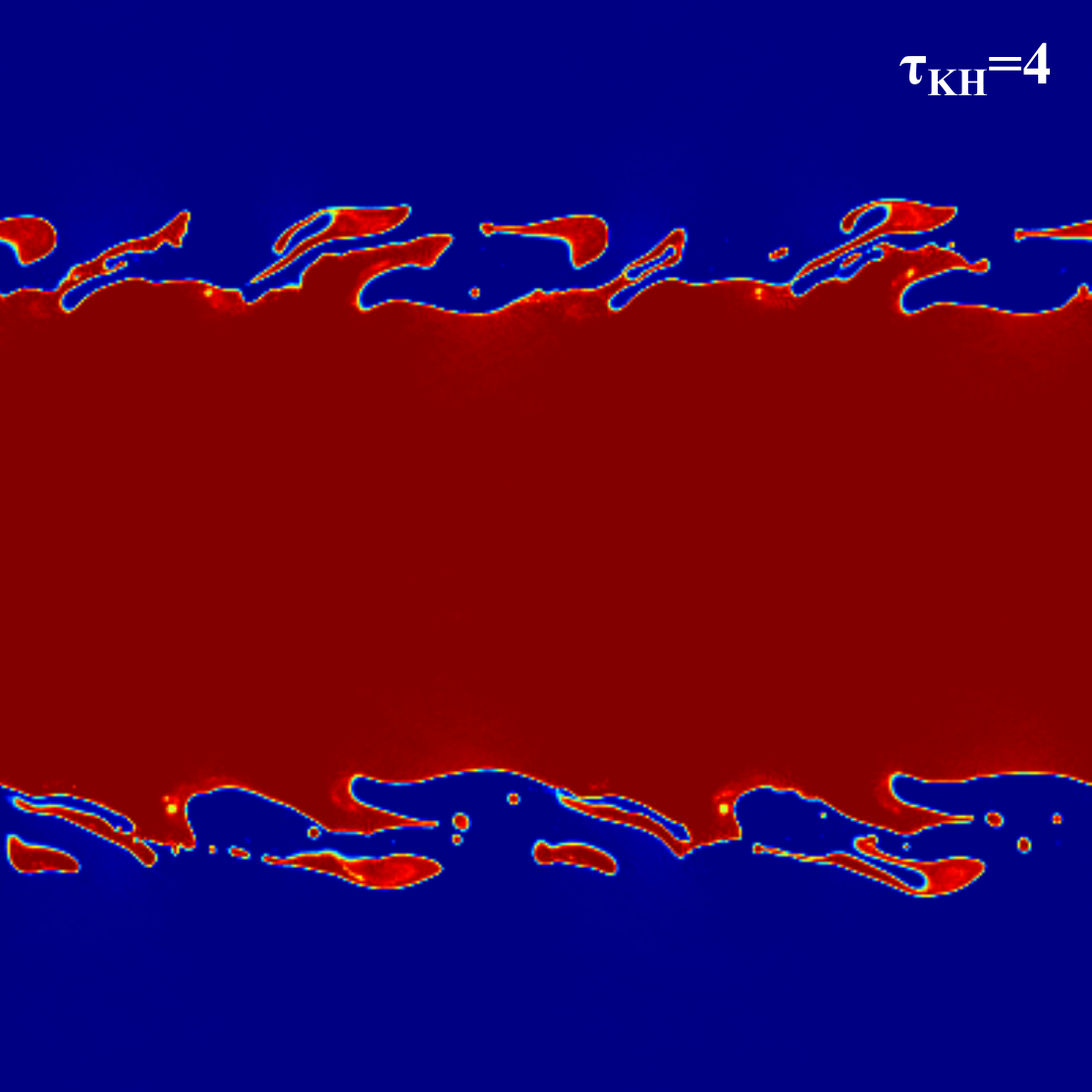}
 & \includegraphics[height=0.2\linewidth]{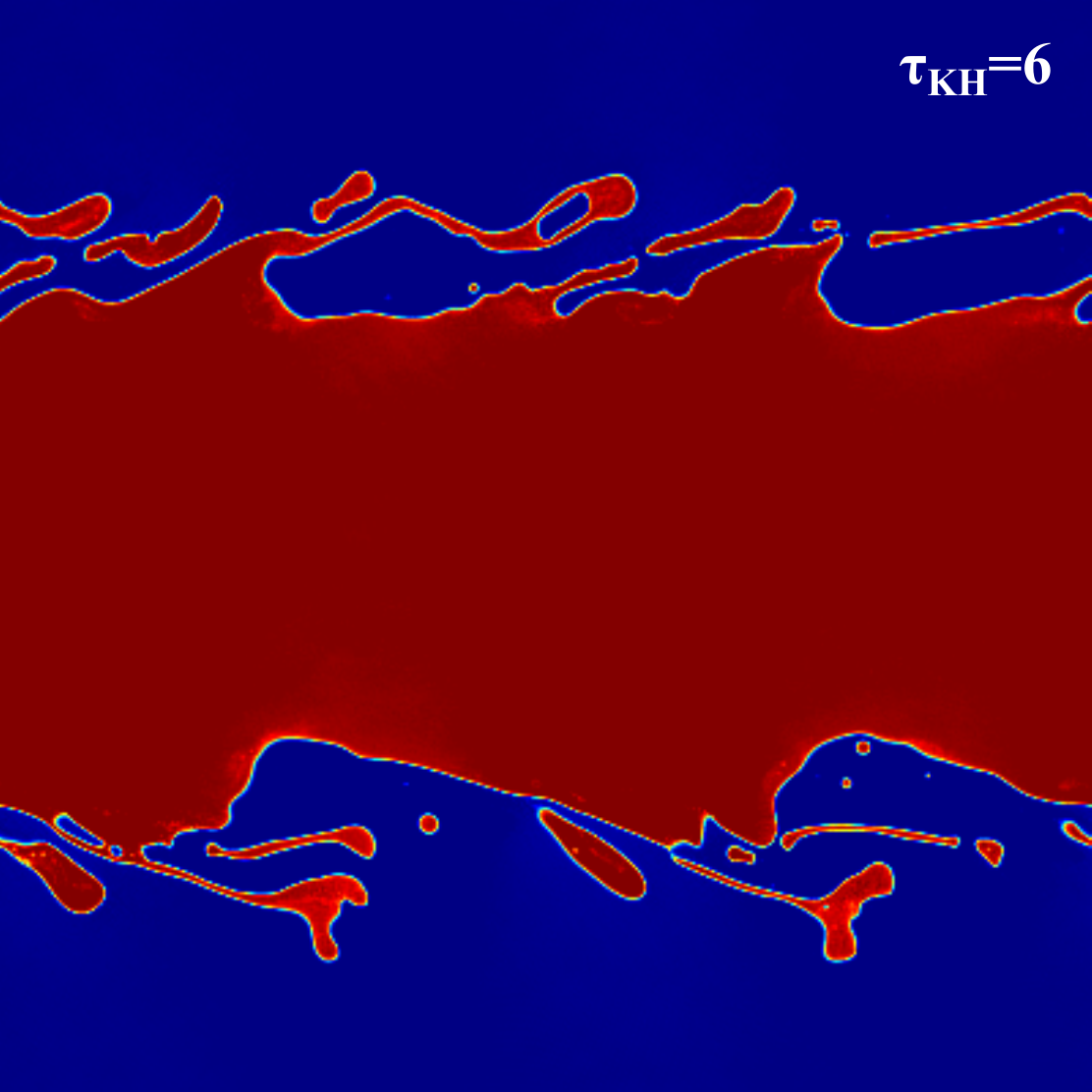}
 & \includegraphics[height=0.2\linewidth]{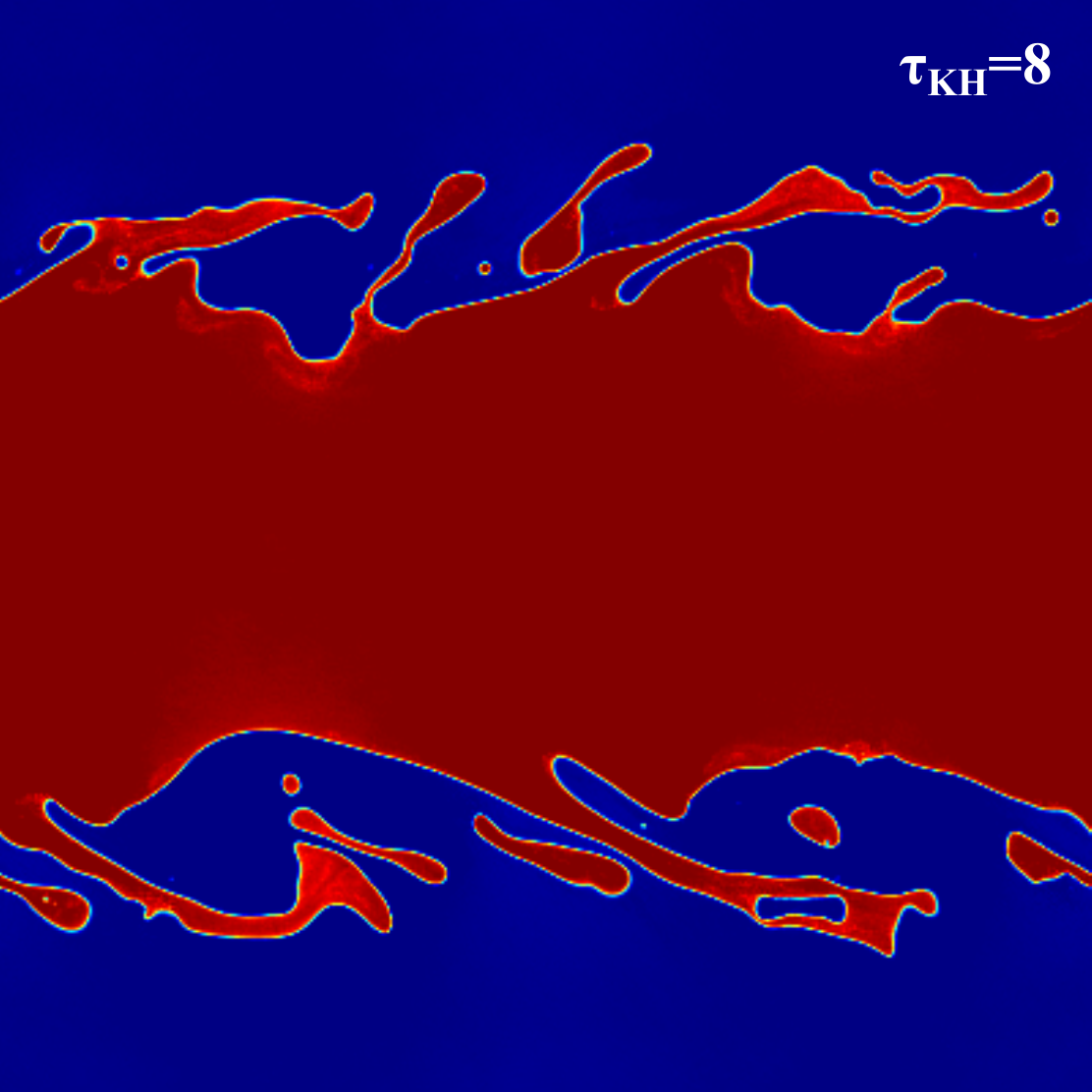}
 & \multirow{3}{*}[0.1836\linewidth]{
\includegraphics[height=0.615\linewidth]{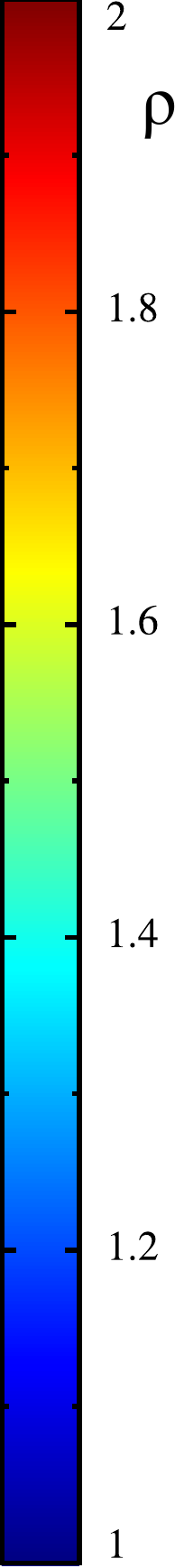}
}
 \\
   \includegraphics[height=0.2\linewidth]{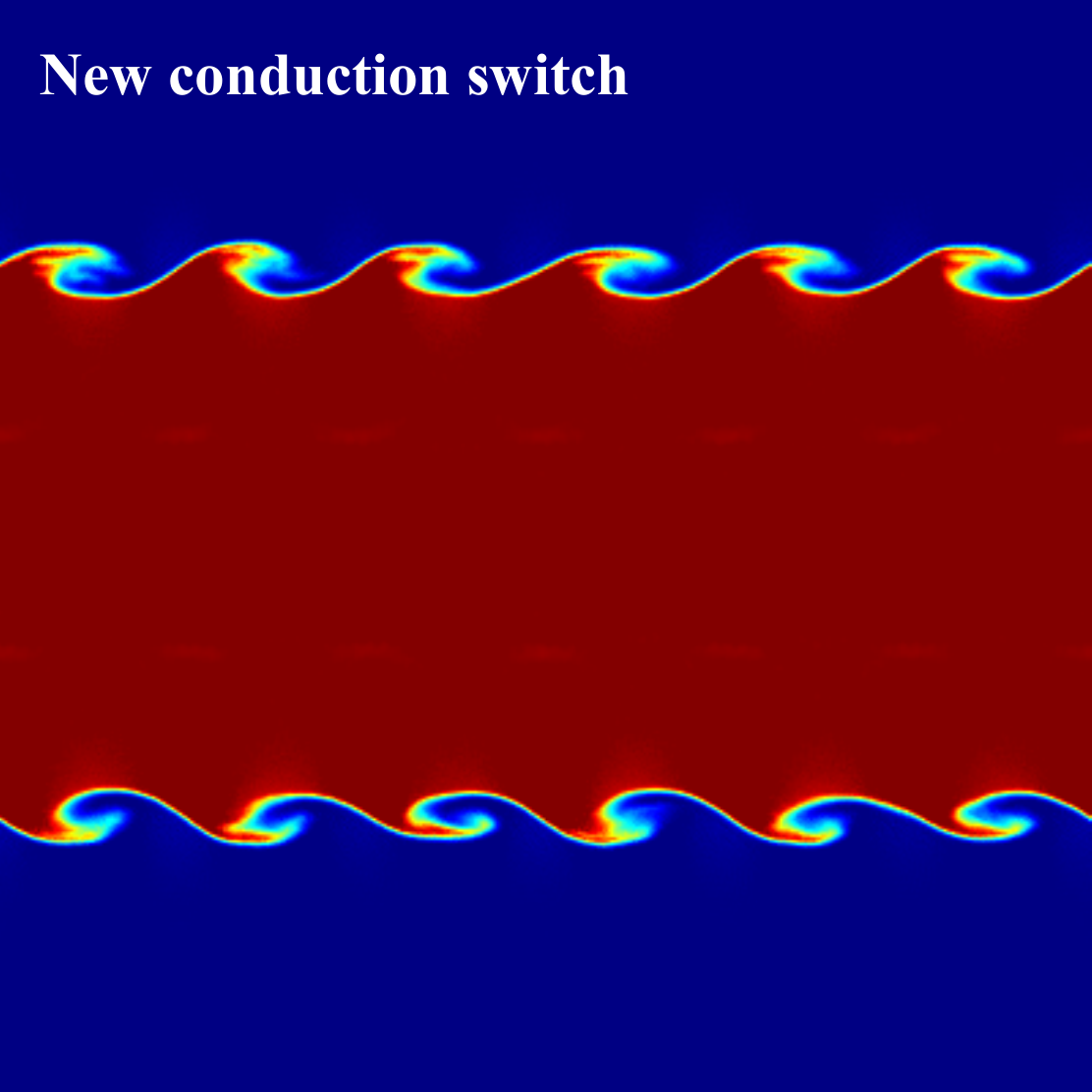}
 & \includegraphics[height=0.2\linewidth]{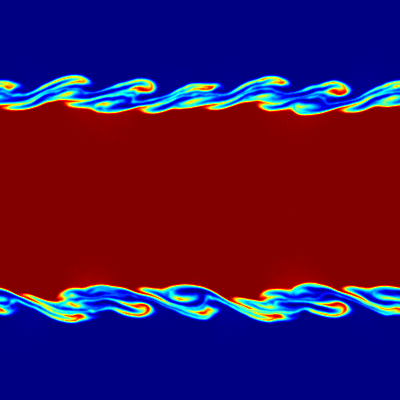}
 & \includegraphics[height=0.2\linewidth]{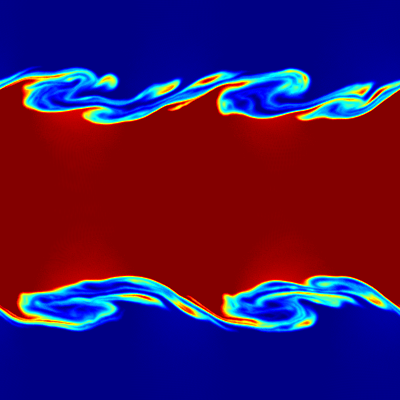}
 & \includegraphics[height=0.2\linewidth]{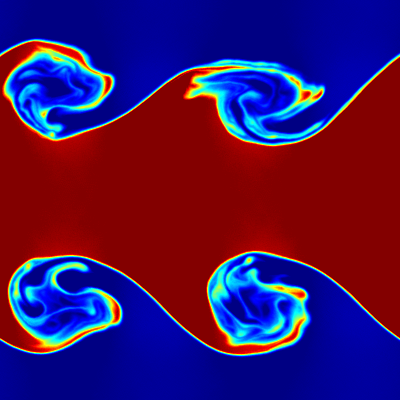}
 &
\\
   \includegraphics[height=0.2\linewidth]{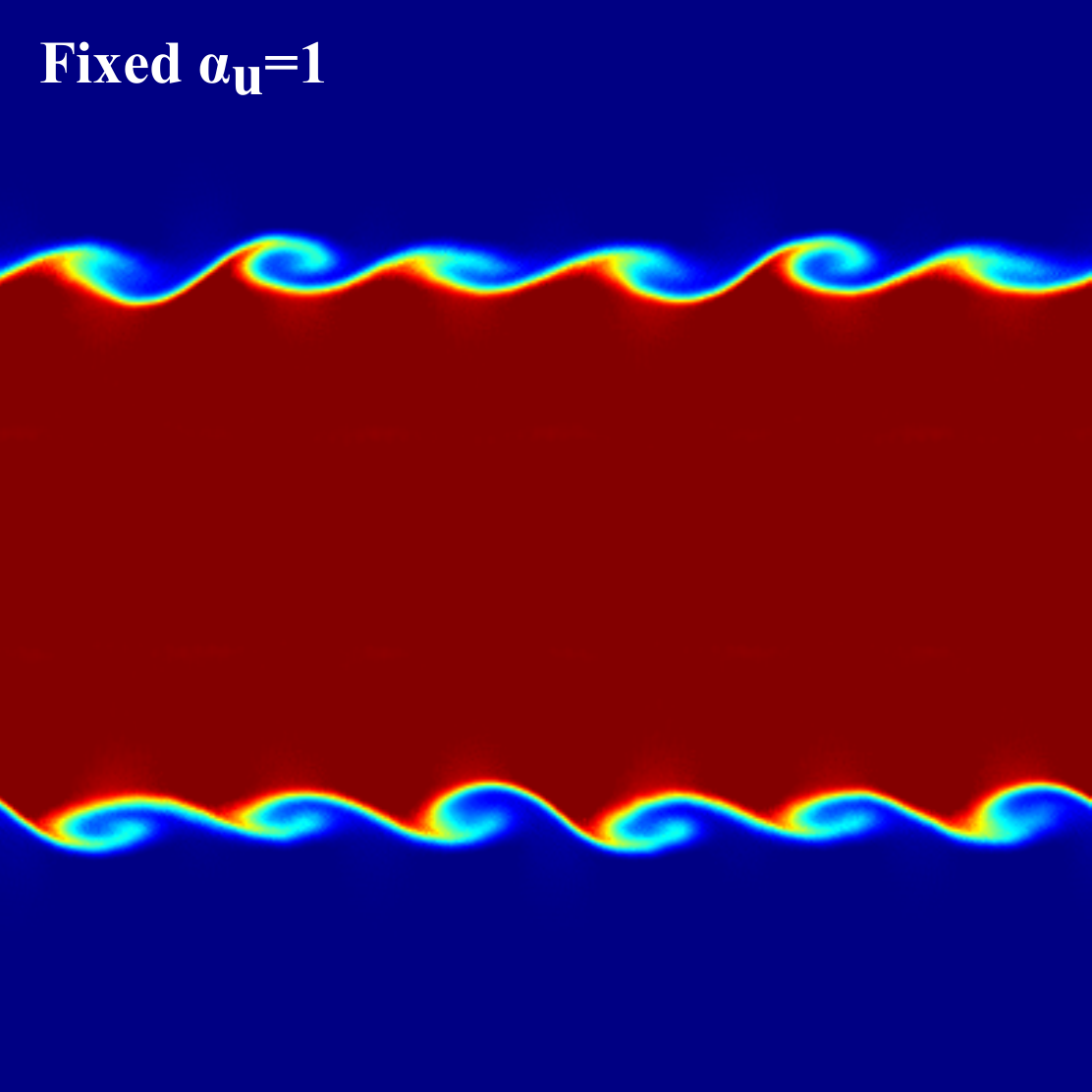}
 & \includegraphics[height=0.2\linewidth]{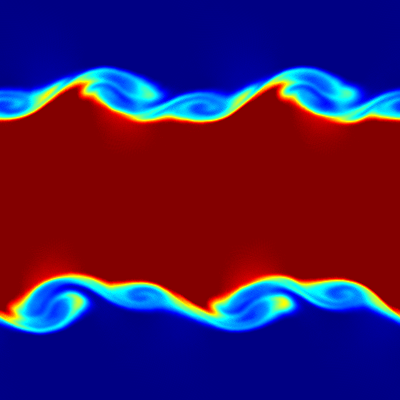}
 & \includegraphics[height=0.2\linewidth]{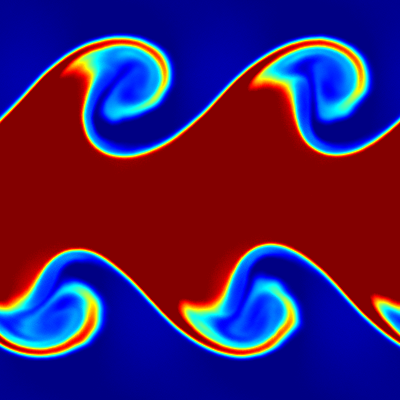}
 & \includegraphics[height=0.2\linewidth]{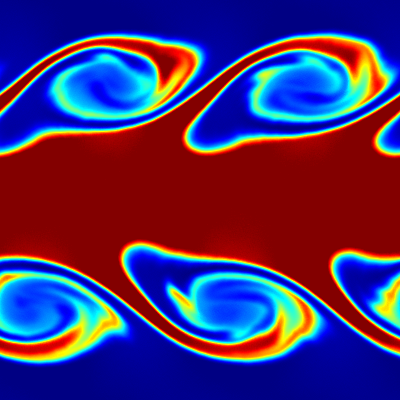}
 &
\end{tabular}
\caption{Results of the Kelvin-Helmholtz instability test using no thermal conduction (top row), the new conduction switch (middle), and thermal conduction with fixed $\alpha_u=1$ (bottom row) for times $\tau_\text{KH}=2,4,6,8$ (left to right).  Instabilities form along the 2:1 density contrast interfaces due to a velocity shear.  Without thermal conduction, the discontinuity in thermal energy is untreated and leads to a spurious pressure preventing the two regions from mixing.  The conduction switch and fixed $\alpha_u=1$ cases smooth the thermal energy discontinuity and the regions mix properly leading to the hallmark billowing curls.}
\label{fig:kh}
\end{figure*}

The particles are initially arranged on triangular lattices.  A total of $454 184$ particles are used, with a particle spacing of $\Delta = 1/512$ in the low density region and $\Delta = 1/724$ in the high density region.  The Morris and Monaghan switch (\ref{eq:intalpha}) is used for artificial viscosity.  The simulation was run with the new thermal conduction switch, and for the limiting cases of no thermal conduction and fixed $\alpha_u=1$ thermal conduction to act as reference comparisons.  Results are presented in Fig.~\ref{fig:kh} at $\tau_\text{KH}=2,4,6,8$.  

The results without thermal conduction demonstrate it's necessity.  The spurious pressure across the interface cause globs of the high density fluid to stretch and break off without mixing and the instability fails to correctly develop.  The new conduction switch however appropriately applies thermal conduction to the interface region, promoting mixing of the fluid.  The results are similar to the fixed $\alpha_u=1$ case, with four large curls being formed at $\tau_\text{KH}=8$.  The interior structure of the curls differ between the two cases, which is to be expected due to the non-linearity of the problem, but the irregular structure of the curls in the switch case suggest more mixing would be beneficial.

\section{Conclusion}
\label{sec:conclusion}

We have developed a new switch for artificial resistivity in Smoothed Particle Magnetohydrodynamics which robustly detects and captures shocks in all field strengths, and also offers reduced dissipation compared to the PM05 switch.  The key design is to measure the relative degree of discontinuity in the magnetic field, setting $\alpha_B = h \vert \nabla {\bf B} \vert / \vert {\bf B} \vert$.  This is simple, yet effective.  By normalising the full gradient of the magnetic field by the magnitude of the magnetic field, the dependence on ${\bf B}$ is removed.  Thus, shocks are able to be captured even in extremely weak fields.  

The switch was tested first for correctness of magnetic shock profiles using a three-dimensional shock tube problem containing three classes of magnetic shocks, yielding excellent agreement to existing solutions.  Comparison with the PM05 switch was performed on the Orszag-Tang vortex problem.  The new switch was found to yield $\alpha_B$ values which closely traced shock lines, and subtle magnetic features were more sharply defined.  The overall dissipation of magnetic energy was significantly reduced, achieving the same effect as if it had been run with increased resolution.  The robustness of the switch was tested by studying driven Mach 10 magnetic turbulence.  The initial magnetic field was extremely weak, with initial magnetic energy 10 orders of magnitude smaller than kinetic energy, yet is strongly shocked due to the turbulence.  Due to the low field strength, the PM05 switch failed to recognise the shocks, leading to their unphysical breakup causing significant noise throughout the magnetic field.  The new switch, howevever, is able to detect and capture the shocks correctly, demonstrating its invariance to field strength.

The design concept of the switch was generalised for use with artificial viscosity and thermal conduction.  The new artificial viscosity switch needed an integrated decay term to treat post-shock oscillations of particle motion, yielding a switch similar to that of Cullen and Dehnen \cite{cd10}.  Performance on a Sod shock tube was in good agreement to the Riemann solution.  The thermal conduction switch was tested with a simulation of the Kelvin-Helmholtz instability.  It was able to mitigate the spurious pressure force across the interface, and the characteristic billowing curls were formed.

Our conclusion is that this artificial resistivity switch is simple, effective, robust, and significantly reduces dissipation of the magnetic field.  In all of these aspects it supercedes the switch of PM05. 

\section*{Acknowledgment}

We thank Christoph Federrath, Guillaume Laibe, and Joe Monaghan for useful discussions.  We appreciate the comments from the referee, Walter Dehnen, on the manuscript of this work submitted to MNRAS.  T. Tricco is supported by Endeavour IPRS and APA postgraduate research scholarships.  We are grateful for funding via Australian Research Council Discovery Projects grant DP1094585.  This research was undertaken with the assistance of resources provided at the Multi-modal Australian ScienceS Imaging and Visualisation Environment (MASSIVE) through the National Computational Merit Allocation Scheme supported by the Australian Government.




%

\bibliographystyle{ieeetr}
\bibliography{sphericbib}

\end{document}